\documentclass[structabstract]{aa}
\usepackage{graphicx}
\usepackage{txfonts}
\begin{document}
   \title{The open cluster NGC6823 and possible triggered star formation associated
with SNR G59.5+0.1}
   \author{Jin-Long Xu
          \inst{1,2,3}
          \ and  Jun-Jie Wang\inst{1,3}}
   \institute{National Astronomical Observatories, Chinese Academy of Sciences,
             Beijing 100012, China \\
              \email{xujl@bao.ac.cn}
         \and
          Graduate University of the Chinese Academy of Sciences, Beijing, 100080, China\\
         \and
             NAOC-TU Joint Center for Astrophysics, Lhasa 850000, China\\
            }
 \authorrunning{J.-L. Xu et al.}
  \titlerunning{NGC6823 and star formation associated with SNR G59.5+0.1}
   \abstract
   {}
   {We investigate the environment in the vicinity of the supernova remnant (SNR) G59.5+0.1 and identify all young stellar
objects (YSOs) around the SNR, to derive the physical properties,
obtain insight into the star-formation history, and further see
whether SNR G59.5+0.1 can trigger star formation in this region.}
   {We have performed the submillimeter/millimeter observations in
CO lines toward the southeast of SNR G59.5+0.1 with the KOSMA 3m
Telescope. High integrated CO line intensity ratio R$_{I_{\rm
CO(3-2)}/I_{\rm CO(2-1)}}$ is identified as one good signature of
SNR-MCs  (molecular clouds) interacting system. To investigate the
impact of SNR G59.5+0.1 on the process of star formation, we used
GLIMPSE I Catalog to select YSOs (including class I and class II
sources).}
{CO emission in an arc-like shape and mid-infrared 8.28 $\mu$m
emission are well coincident with  SNR G59.5+0.1, which has the
total mass of $1.1\times10^{4}\rm~M_{\odot}$ and fully cover open
cluster NGC 6823. Three molecular clumps were identified in the CO
molecular arc, each clump shows the broad line wing emission,
indicating that there are three outflows motion. The integrated CO
line intensity ratio (R$_{I_{\rm CO(3-2)}/I_{\rm CO(2-1)}}$) for the
whole molecular arc is between 0.48 and 1.57. The maximum value is
1.57, which is much higher than previous measurements of individual
Galactic MCs. The CO molecular arc has a line intensity ratio
gradient. SNR G59.5+0.1 is in adiabatic expansion phase. The age of
the SNR is 8.6 $\times$ $10^{4}$ yr. Based on GLIMPSE I Catalog 625
young stellar objects (YSOs) candidates (including 176 class I
sources and 449 class II sources) are selected. The timescales for
class 0, class I and class II sources are $\leq10^{4}$ yr,
$\sim10^{5}$ yr, and $\sim10^{6}$ yr, respectively. The number of
YSOs are significantly enhanced in the interacting regions,
indicating the presence of some recently formed stars. }
   {}


   \maketitle
%

\section{Introduction}
Massive star has significant impact on the morphology and chemical
evolution of the surrounding Interstellar Medium (ISM) through
UV-radiation, stellar winds, and supernova explosion. Stars form
from the densest environments in molecular clouds (MCs), often
clustered together spatially in groups ranging from a few sources to
many thousands. When a supernova explodes near MCs, shock generated
by supernova remnant (SNR) may compress some condensed MCs to
collapse. On the other hand, the shock can enhance abundances of the
different molecular species with respect to quiescent cloud
conditions. So the SNR-MCs interaction plays an essential part in
the  processes of star formation and the evolution of ISM (Reynoso
et al. \cite{Reynoso}).  These associations are established by the
detection of OH 1720 MHz masers, the broad emission lines, high
ratio of $^{12}$CO $J=2-1 $ to $J=1-0 $ integrated line intensity,
$\gamma$-ray emission from MCs toward SNR as well as the
morphological relation between SNR and MCs, and so on (Frail et al.
\cite{Frail}; Green et al. \cite{Green}; Seta et al. \cite{Seta};
Wilner et al. \cite{Wilner}; Huang \& Thaddeus \cite{Huang}; Byun et
al. \cite{Byun}; Hewitt \& Yusef-Zadeh \cite{Hewitt}). Jiang et al
(\cite{Jiang}) summarized the criteria of the SNR-MCs association.
In addition, some results from numerical simulation have speculated
that star formation can be triggered by the SNR-MCs interaction
(Melioli et al. \cite{Melioli} \&  Leao et al. \cite{Leao}).
However, only a few cases about the triggered star formation have
been proposed from the observed results (Junkes et al.
\cite{Junkes}; Parons et al. \cite{Parons}; Xu et al. \cite{xua}) .

SNR G59.5+0.1, at a distance of 2.1$\sim$2.3 kpc (Xu et al.
\cite{xuj}; Billot et al. \cite{billot}), is situated in the
direction of Vulpecula OB at the position ($l, b$) = (59.58, 0.12).
According to Xu et al. (\cite{xuj}), the age of SNR G59.5+0.1 is
ranges from 10$^{3}$ to 10$^{4}$ years old. Taylor et al.
(\cite{Taylor}) described this object as a shell-type SNR with
nonthermal spectral index ($\alpha$ $<$ -0.4) and with a diameter of
15$^{\prime}$. They also mentioned that the SNR is possibly
associated with the HII region Sh2-86. Billot et al. (\cite{billot})
found a circular structure centered at the position of the SNR with
a diameter of 15$^{\prime}$, which confirms the value found by
Taylor et al. (\cite{Taylor}). Furthermore, Billot et al.
(\cite{billot}) noted a high density of YSOs lined up along two arcs
at the northern and southern rims of SNR G59.5+0.1, suggesting that
the interaction of the expanding shell of SNR G59.5+0.1 and the
neutral gas of Sh2-86 could have triggered the formation of young
stars in the cluster. NGC6823 is an open cluster located at
R.A(J2000)=$19^{\rm h}43^{\rm m}08^{\rm s}.4$,
Dec(J2000)=$23^{\circ}18'00^{\prime\prime}$ with a diameter of
$\sim$ 16$^{\prime}$ (Kharchenko et al. \cite{Kharchenko}). Riaz et
al. (\cite{Riaz}) suggested that NGC6823 have possibly experienced a
recent burst of star formation, which may be caused by a supernova
explosion of massive O star. However, they suggested that SNR
G59.5+0.1 is not old enough to trigger any of their identified star
formation.

Motivated by the supposed association of SNR G59.5+0.1 with
molecular gas and with possibly triggered star formation, we have
performed $^{12}$CO $J=2-1 $, $^{12}$CO $J=3-2$, and
$^{13}$CO$J=3-2$ observations toward G59.5+0.1. The observations are
described in $\S$2, and the results are presented in $\S$3. In
$\S$4, we discuss how our data lend supportive evidence of the
triggered star formation in the interacting region. The conclusions
are summarized in $\S$5.

  \begin{figure}
   \centering
  \includegraphics[angle=0,scale=.4]{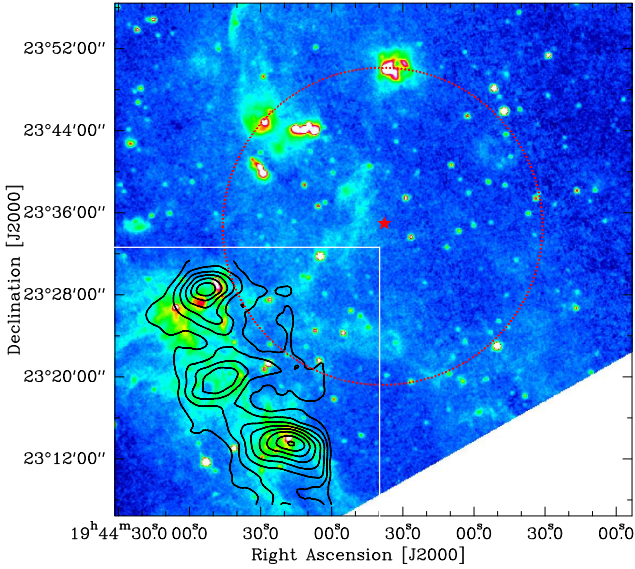}
   \caption{ $^{12}$CO $J=2-1$ intensity map (black
contours) integrated from 18 to 35 km s$^{-1}$, overlayed on the the
mid-infrared 8.28-$\mu$m MSX emission map (color scale). The contour
levels are 30, 40, ... , $90\%$ of the peak value (83.9 K km
s$^{-1}$). The red star represents the center of SNR G59.5+0.1 and
its extent is outlined by a red dashed circle. The radius of SNR
G59.5+0.1 is 15$^{\prime}$ (Taylor et al. 1992 \& Billot et al.
2010). The white box outlines the field seen on the Fig.2. }
\end{figure}

   \begin{figure}
\vspace{0mm}
   \centering
   \includegraphics[angle=270,scale=.6]{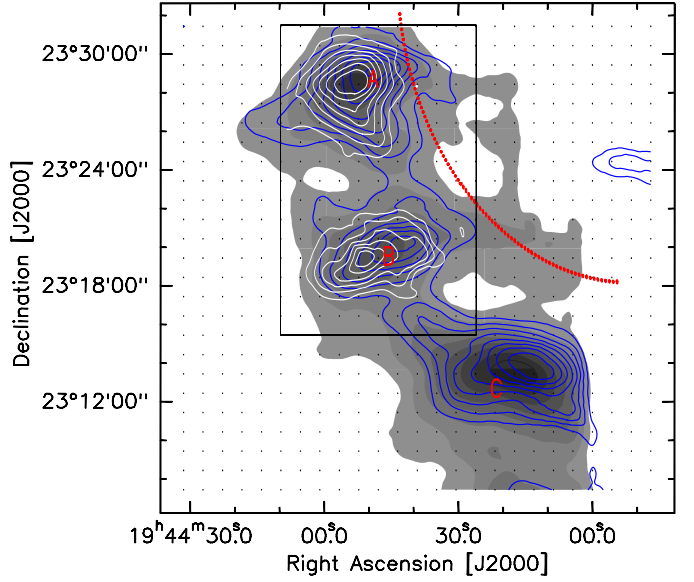}
 \vspace{0mm} \caption{$^{12}$CO $J=2-1$ intensity map (gray scale), overlayed on $^{12}$CO $J=3-2$ (blue contours) and $^{13}$CO $J=2-1$ (white contours) intensity maps.
The contour levels of each CO molecule are 30, 40,..., $90\%$ of the
peak value.  The $^{12}$CO $J=3-2$ and $^{13}$CO $J=2-1$ peak values
are 79.0 and 15.5 K km s$^{-1}$, respectively. Letter A, B and C
indicate the different MC clumps. The dot symbols mark the mapping
points of $^{12}$CO $J=2-1$ and $^{12}$CO $J=3-2$, while the black
rectangle outlines the mapped field of $^{13}$CO $J=2-1$.}
   \end{figure}


\begin{figure*}
\vspace{-8mm} \centering
\includegraphics[angle=270,scale=.66]{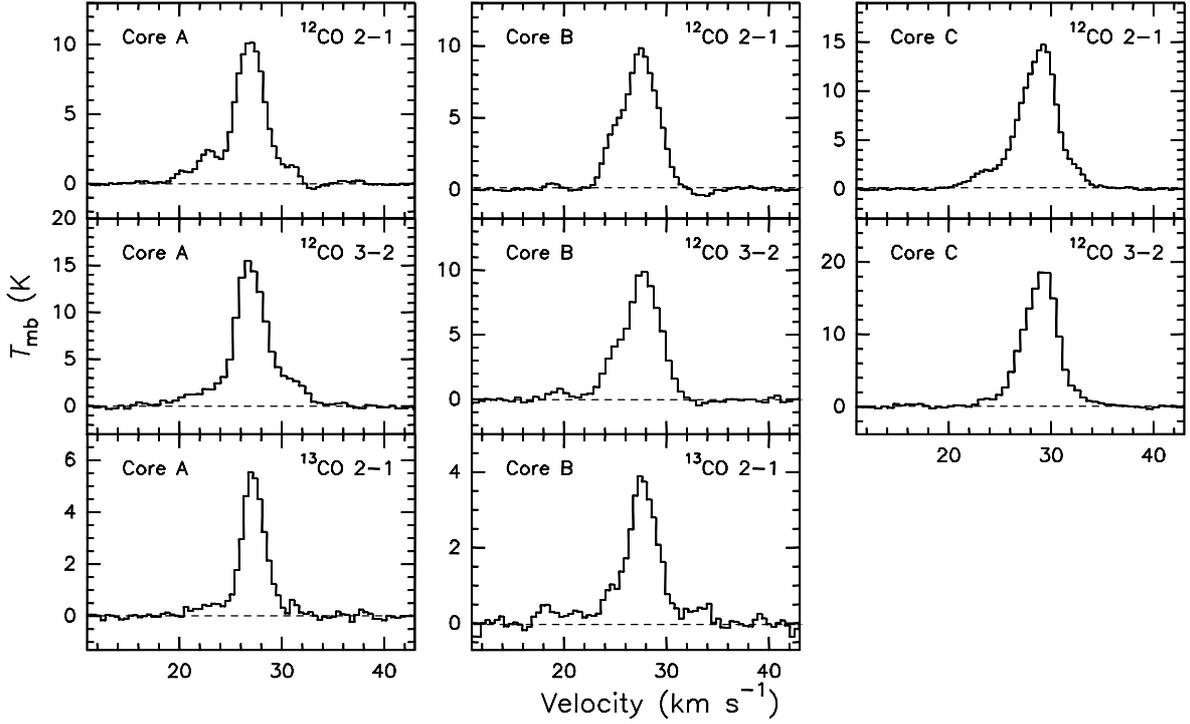}
\vspace{-14mm} \caption{Averaged spectra of molecular line $^{12}$CO
$J=2-1$, $^{12}$CO $J=3-2$ and $^{13}$CO $J=2-1$ over each clump.
The dashed lines mark the position at the zero intensity.}
\end{figure*}


\begin{table*}[]
\begin{center}
\tabcolsep 3.2mm\caption{Observed parameters of each line}
\def\temptablewidth{10\textwidth}%
\begin{tabular}{lcccccccccc}
\hline\hline\noalign{\smallskip}
Name   &      & $^{12}$CO 2-1   &   &   & $^{12}$CO 3-2 &  & & $^{13}$CO 2-1& \\
\cline{2-10}
        &   $T_{\rm mb}$   &FWHM  &$V_{\rm LSR}$   &   $T_{\rm mb}$   &FWHM  &$V_{\rm LSR}$&   $T_{\rm mb}$   &FWHM  &$V_{\rm LSR}$   \\
        &  (K)               &(km $\rm s^{-1}$) &(km $\rm s^{-1}$) &   (K)   &(km $\rm s^{-1}$)  &(km $\rm s^{-1}$)&   (K)   &(km $\rm s^{-1}$)&(km $\rm s^{-1}$) \\
\hline\noalign{\smallskip}
Core A  & 9.9  (0.6) & 3.7 (0.2)  & 27.0  (0.1) &  14.5 (0.4)  & 3.9(0.1) & 27.0 (0.1)  & 5.4 (0.2) & 2.7 (0.1) & 27.3 (0.1) \\  
Core B  & 9.2  (0.2) & 4.4 (0.1)  & 27.4 (0.1)  &  9.4  (0.2)  & 4.4(0.1) & 27.6 (0.1)  & 3.6 (0.2) & 3.6 (0.2) & 27.6 (0.2) \\  
Core C  & 13.9 (0.3) & 4.4 (0.1)  & 28.8 (0.1)  &  18.3 (0.3)  & 4.0(0.1) & 29.0 (0.1)  & --- & --- &--- \\  
\noalign{\smallskip}\hline
\end{tabular}\end{center}
Note: The position of clump A is at RA(J2000)=$19^{\rm h}43^{\rm
m}51.85^{\rm s}$ and DEC(J2000)=$23^{\circ}28'26.34^{\prime\prime}$;
The position of clump B is at RA(J2000)=$19^{\rm h}43^{\rm
m}47.85^{\rm s}$ and DEC(J2000)=$23^{\circ}19'26.34^{\prime\prime}$;
The position of clump C is at RA(J2000)=$19^{\rm h}43^{\rm
m}15.85^{\rm s}$ and DEC(J2000)=$23^{\circ}13'26.34^{\prime\prime}$.
\end{table*}


\section{Observations}
The mapping observations of G59.5+0.1 were made in $^{12}$CO $J=2-1
$, $^{12}$CO $J=3-2$ and $^{13}$CO $J=2-1$ lines using the KOSMA 3m
telescope at Gornergrat, Switzerland,  in April 2004. The half-power
beam widths of the telescope at the observing frequencies of 230.538
GHz, 345.789 GHz and 220.399 GHz are $130^{\prime\prime}$,
$80^{\prime\prime}$, and $130^{\prime\prime}$, respectively. The
pointing and tracking accuracy is better than 20$^{\prime\prime}$.
The accuracy of the absolute intensity calibration to be better than
15\% (Sun et al. \cite{Sun}). The DSB receiver noise temperature was
about 170 K. The medium and variable resolution acousto optical
spectrometers have 1501 and 1601 channels, with total bandwidth of
248 MHz and 544 MHz. The channel widths of 165 and 340 kHz
correspond to velocity resolutions of 0.21 and 0.29 ${\rm km\
s^{-1}}$, respectively. The beam efficiency $B_{\rm eff}$ is 0.68 at
230 GHz and 220 GHz. The beam efficiency $B_{\rm eff}$ is 0.72 at
345 GHz. The forward efficiency $F_{\rm eff}$ is 0.93.  Mapping
observations are centered at RA(J2000)=$19^{\rm h}43^{\rm
m}47.85^{\rm s}$, DEC(J2000)=$23^{\circ}21'26.34^{\prime\prime}$
using the On-The-Fly mode, the total mapping area is
$24^{\prime}\times 24^{\prime}$ in
 $^{12}$CO $J=2-1 $ and $^{12}$CO $J=3-2 $ with a $1^{\prime}\times 1^{\prime}$
 grid, while the mapping area in $^{13}$CO $J=2-1 $ is $10^{\prime}\times
 15^{\prime}$.

The data were reduced using the GILDAS/CLASS
\footnote{http://www.iram.fr/IRAMFR/GILDAS/} package. The correction
for the line intensities to {the }main beam temperature scale was
made using the formula $T_{\rm mb}= (F_{\rm eff}/B_{\rm eff}\times
T^\ast_{\rm A})$.

\section{Results}
\subsection{Molecular emission}
Figure 1 presents the emission map (color scale) at 8.28 $\mu$m. The
emission shows a half-shell morphology, which is associated with SNR
G59.5+0.1 with a diameter of 15$^{\prime}$. The integrated intensity
map of $^{12}$CO $J=2-1$ is overlapped the 8.28 $\mu$m emission
located in the white boxed area of Figure 1, which covers open
cluster NGC6823 marked in a green dashed circle (see Figure 8). The
$^{12}$CO $J=2-1 $ emission coincides well with the 8.28 $\mu$m
emission at the southeast. In order to further analyze the
morphology of CO molecular gas, we make the integrated intensity map
of $^{12}$CO $J=3-2$ and $^{13}$CO $J=2-1 $ (Figure 2). In Figure 2,
$^{12}$CO $J=2-1 $ emission is coincident with that from $^{12}$CO
$J=3-2$ and $^{13}$CO $J=2-1 $. The $^{12}$CO $J=2-1 $ and $^{12}$CO
$J=3-2 $ emission appear to show a arc-like structure, while the
$^{13}$CO $J=2-1 $ emission may trace the compact core of molecular
gas. We find three cloud clumps in this structure, each clump is
designated alphabetically, clump A, clump B and  clump C. Because
the mapping area in $^{13}$CO $J=2-1$ is smaller (see section 2), we
did not detected the $^{13}$CO $J=2-1 $ emission from the clump C.
Figure 3 shows $^{12}$CO $J=2-1$, $^{12}$CO $J=3-2$ and $^{13}$CO
$J=2-1$ spectra averaged over clump A, clump B and clump C,
respectively. Line profiles of $^{12}$CO $J=2-1$ and $^{12}$CO
$J=3-2$ for each clump appear to be broadened, the velocity
components are mainly located in interval 18 to 35 km s$^{-1}$. The
spectral profile of clump A may show a broad outflow wing and has
two weaker components at 23 km s$^{-1}$ and 31 km s$^{-1}$. We have
made Gaussian fits to all the spectra. The fitted results are
summarized in Table 1. From Table 1, we derive systemic velocities
of $\sim$27.1 km s$^{-1}$, $\sim$27.5 km s$^{-1}$, and $\sim$28.9 km
s$^{-1}$ in clump A, clump B, and clump C, respectively. According
to the galactic rotation model of Fich et al. (\cite{Fich}) together
with $R_{\odot}$ = 8.5 kpc and $V_{\odot}$ = 220 km s$^{-1}$, where
$V_{\odot}$ is the circular rotation speed of the Galaxy, we obtain
a kinematic of ~2.3 kpc for the distance of all clumps, which is
roughly consistent with the photometric distance (2.1$\pm$0.1 kpc)
of NGC 6823 (Guetter \cite{guetter}).

Assuming local thermodynamical equilibrium (LTE) and using the
$^{12}$CO $J=2-1$ line,  the column density of the clumps in
cm$^{-2}$ is estimated as (Garden et al. \cite{Garden}; Xu et al.
\cite{xu})
\begin{equation}
\mathit{N_{\rm clu}}=1.08\times10^{13}\frac{(T_{\rm
ex}+0.92)}{\exp(-16.62/T_{\rm ex})}\int T_{\rm mb}\times\frac{\tau
\rm dv}{[1-\exp(-\tau)]},
\end{equation}\indent
where $\rm dv$ is the velocity range in km s$^{-1}$, $T_{\rm ex}$ is
the excitation temperature in K, and $\tau$ is the optical depth for
$^{12}$CO $J=2-1$ line. $T_{\rm ex}$ is estimated following the
equation $T_{\rm ex}=11.1/{\ln[1+1/(T_{\rm mb}/11.1+0.02)]}$, while
we calculate $\tau$ according to

\begin{equation}
\frac{T_{\rm mb}(^{12}\rm CO)}{T_{\rm mb}(^{13}\rm
CO)}\approx\frac{1-\rm exp(-\tau)}{1-\rm exp(-\tau/89)},
\end{equation}

where $T_{\rm mb}$ is the corrected main beam temperature.  Here we
assume the solar abundance ratio of $[^{12}\rm CO]/[^{13}\rm CO]$=89
(Lang \cite{Lang}; Anders \& Grevesse \cite{Andres}; Garden et al.
\cite{Garden}), and that the $^{12}$CO $J=2-1$ emission is optically
thick. Additionally, we use the relation $N_{\rm H_{2}}$ $\approx$
$10^{4}N_{\rm ^{12}CO}$ (Dickman \cite{Dickman}). If the clumps are
approximately spherical in shape, the mean number density $\rm
H_{2}$ is $n(\rm H_{2})$=$1.62\times10^{-19}N_{\rm H_{2}}/L$, where
$L$ is the clump diameter in parsecs (pc). Furthermore, their mass
is given by $M_{\rm H_{2}}$=$n(\rm H_{2})$$\frac{1}{6}\pi
L^{3}\mu_{g}m(\rm H_{2})$ (Garden et al. \cite{Garden}), where
$\mu_{g}$=1.36 is the mean atomic weight of the gas, and $m(\rm
H_{2})$ is the mass of a hydrogen molecule. The derived column
density, mean number density, and mass of each clump may be in error
by as much as a factor of three depending on the exact value of the
$[^{12}\rm CO]/[^{13}\rm CO]$ abundance ratio (Garden et al.
\cite{Garden}), which are all listed in Table 2.

\begin{table}[]
\begin{center}
\tabcolsep 2.5mm \caption{The physical parameters of the clumps in
LTE.}
\def\temptablewidth{0.1\textwidth}
\begin{tabular}{ccccccccc}
\hline\hline\noalign{\smallskip}
Name   &$T_{\rm ex}$& $L$ &$N_{\rm H_{2}}$ & $n(\rm H_{2})$& $M$   \\
       & K&  pc  &(cm$^{-2}$) &(cm$^{-3}$)  & ($10^{3}\rm M_{ \odot}$)     \\
  \hline\noalign{\smallskip}
Clump A  &15.0 &  5.4 & 1.4$\times10^{22}$  & 4.2$\times10^{2}$ & 6.3   \\  
Clump B   &14.3&  4.7 & 0.9$\times10^{22}$  & 3.3$\times10^{2}$ & 3.2  \\  
Clump C  & 19.2&  6.3 & 2.2$\times10^{22}$  & 5.6$\times10^{2}$ & 1.4  \\  
\noalign{\smallskip}\hline
\end{tabular}\end{center}
\end{table}

   \begin{figure*}
   \vspace{0mm}
   \centering
   \includegraphics[angle=270,scale=.8]{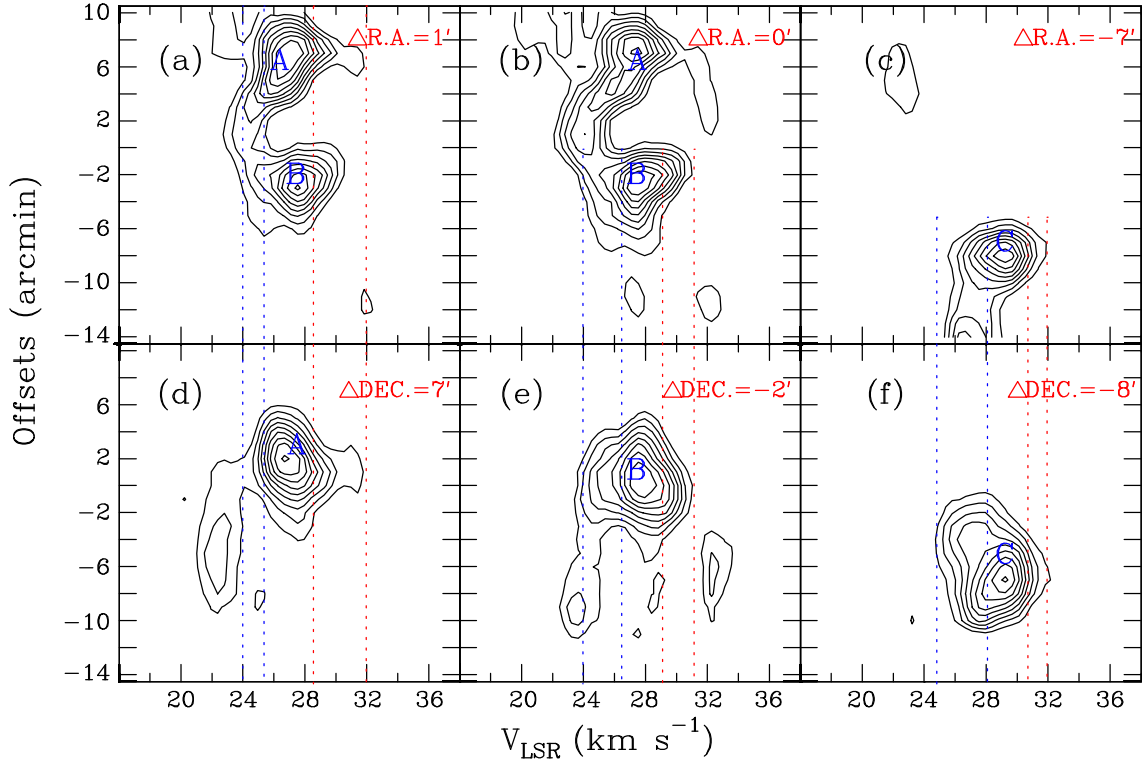}
\vspace{0mm} \caption{ P-V diagram constructed from  CO $J=2-1$
transition for clump A-C. (a)-(c) panels: Contour levels are 10,
20,..., $90\%$ of the peak value, with a cut through the peak
positions of each clump along the north-south direction, the peak
positions are shown in the upper right corner. (d)-(f) panels:
Contour levels are the same as above panels, with a cut along the
east-west direction. The blue and red vertical dashed lines mark the
velocity ranges of the blueshifted and redshifted emission for each
clump, respectively. The offsets given in each panel are relative to
RA(J2000)=$19^{\rm h}43^{\rm m}47.85^{\rm s}$ and
DEC(J2000)=$23^{\circ}21'26.34^{\prime\prime}$.}
\end{figure*}

To determine the velocity components and confirm the outflow, we
have made position-velocity (PV) diagrams with the cuts through the
peak positions of each clump along the north-south and east-west
directions, as shown in Figure 4. From Figure 4, we can clearly
identify three clumps. Each clump shows the bipolar structure marked
by the blue and red vertical dashed lines.  The blueshifted and
redshifted components have obvious velocity gradients, in particular
for clump A and clump B presented in panel (d) and (e) of Figure 4.
The distributions of redshifted and blueshifted velocity components
are a indication of outflow motion. In panel (d), we also identify a
velocity component from 21 to 24 km $\rm s^{-1}$, which is not from
the component of the outflow, but related to a $^{12}$CO $J=2-1 $
weaker component peaked at -23 km $\rm s^{-1}$ as seen from the
$^{12}$CO $J=2-1 $ spectra of clump A in Figure 3. Using the
velocity ranges obtained from the PV diagram, we made the velocity
integrated intensity maps superimposed on each clump emission map,
respectively. The distribution of redshifted and blueshifted
velocity components in Figure 5 provides us further evidence for the
bipolar outflow in each clump.

   \begin{figure}
   \vspace{0mm}
   \centering
   \includegraphics[angle=270,scale=.42]{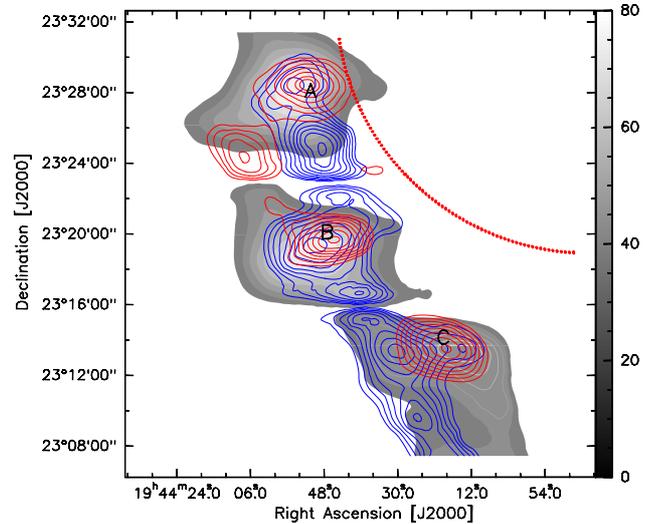}
\vspace{0mm} \caption{ The velocity integrated intensity maps of
$^{12}$CO $J=2-1$ outflows (red and blue contours) overlaid with the
$^{12}$CO $J=2-1$ emission of each clump (gray scale). The red and
blue contour levels are 20, 30,...,100\% of the peak value. The red
dashed arc represents SNR G59.5+0.1.}
\end{figure}
   \begin{figure}
   \centering
   \includegraphics[angle=270,scale=.55]{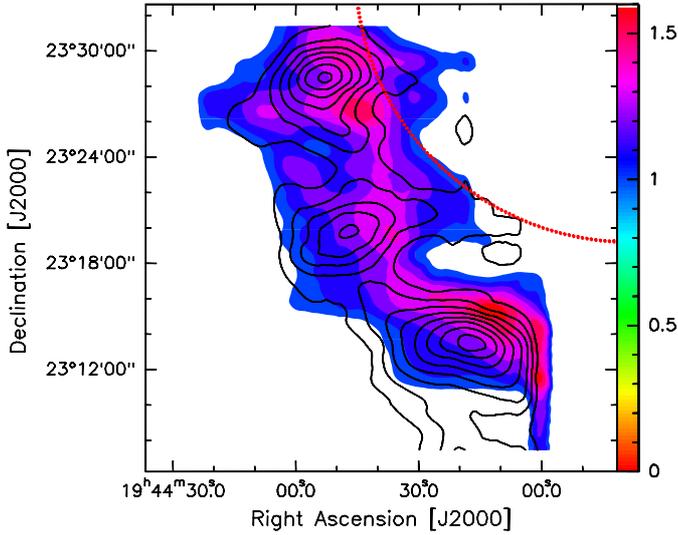}
      \caption{$^{12}$CO $J=2-1$ intensity map are superimposed on the line
intensity ratio map (color scale), the line intensity ratios
(R$_{I_{\rm CO(3-2)}/I_{\rm CO(2-1)}})$ range from 0.48 to 1.57 by
0.16. The wedge indicates the line intensity ratios scale.}
\end{figure}

When a supernova expands into the surrounding molecular clouds, the
shocks resulting from the interaction of SNR with MCs can heat the
surrounding gases. As the temperature of gases increases, the line
opacities decrease as the upper J levels become more populated, then
the integrated intensity ratio of $^{12}$CO $J=3-2 $ to $^{12}$CO
$J=2-1 $ (R$_{I_{\rm CO(3-2)}/I_{\rm CO(2-1)}}$) can indicate the
shock (Xu et al. \cite{xua}; \cite{xub}). In order to obtain the
integrated intensity ratio of $^{12}$CO $J=3-2 $ to $^{12}$CO $J=2-1
$,  the $80^{\prime\prime}$ resolution of $^{12}$CO $J=3-2$ data is
convolved with an effective beam size of
$\sqrt{130^{2}-80^{2}}=102^{\prime\prime}$. We calculated the
integrated intensities  for $^{12}$CO $J=3-2$ line in the same
velocity range as for $^{12}$CO $J=2-1$ line. The integrated range
is from 16 to 36 km s$^{-1}$. Figure 6 shows the distribution of
R$_{I_{\rm CO(3-2)}/I_{\rm CO(2-1)}}$ (color scale) overlaid with
the distribution of $^{12}$CO $J=2-1$ line integrated intensity
(contours). The red dashed arc represents SNR G59.5+0.1. In
Figure.6, the whole arc-like molecular gas has a ratio value
gradient increasing from northeast to southwest, suggesting that SNR
shock is expanding into clump A-C and start to compress each clump.
The ratio values for the clump A, clump B, and clump C are between
0.48 and 1.57. The maximum value is 1.57, which is much higher than
typical value (0.55) for MCs in the Galactic disk (Sanders et al.
\cite{sanders}) and value (0.69) for the normal MCs in M33 (0.69
Wilson et al. \cite{wilson}), and even higher than value (0.8) in
the starburst galaxies M82 (Guesten et al. \cite{guesten}).

   \begin{figure}
  \vspace{-9.5mm}
   \centering
   \includegraphics[angle=0,scale=.85]{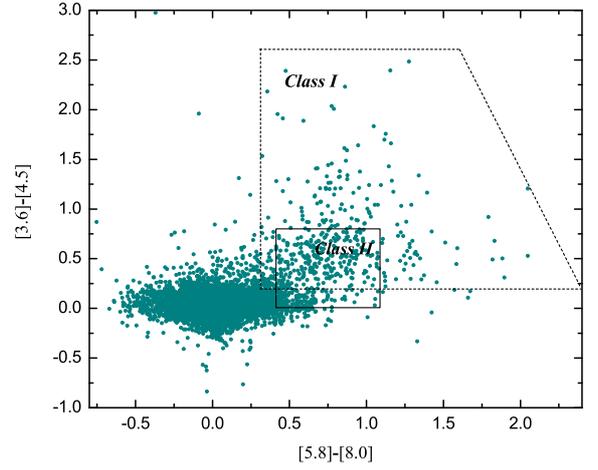}
  \vspace{-7mm}
      \caption{GLIMPSE color--color diagram [5.8]--[8.0] versus [3.6]--[4.5]
for sources. The regions indicate the stellar evolutionary stage as
defined by Allen et al. (2004). Class I sources are protostar with
circumstellar envelopes and class II are disk dominated objects.}
   \end{figure}

\subsection{Infrared emission}
To further search for primary tracers of the star-formation activity
in the surrounding of G59.5+0.1, we used the GLIMPSE I Catalog which
consists of point sources that are detected at least twice in one
band. The GLIMPSE I survey observed the Galactic plane (65$^{\circ}$
$<$ $|l|$ $<$ 10$^{\circ}$ for $|b|$ $<$ 1$^{\circ}$) with the four
mid-IR bands (3.6, 4.5, 5.8, and 8.0 $\mu$m) of the Infrared Array
Camera (IRAC; Fazio et al. \cite{Fazio}) on the Spitzer Space
Telescope. From the database, we  selected 29469 near-infrared
sources within a circle of 24$^{\prime}$ in radius centered on
R.A.=19$^{\rm h}42^{\rm m}50^{\rm s}$ (J2000),
Dec=$+23^{\circ}30'00^{\prime\prime}$ (J2000). The size of this
region completely covers the extension of G59.5+0.1 and the open
cluster NGC6823.  Figure 7 shows the $[5.8]-[8.0]$ versus
$[3.6]-[4.5]$ color-color (CC) diagram. The regions in the figure
indicate the stellar evolutionary stages based on the criteria of
Allen et al. (\cite{Allen}),  Parons et al (\cite{Parons}), and
Petriella et al (\cite{Petriella10a}).  The near-infrared sources
are classified into three regions: class I sources are protostars
with circumstellar envelopes (polygon), class II sources are
disk-dominated objects (rectangle), and other sources.  Using this
criteria, we find 176 class I sources and 449 class II sources. Here
class I and class II sources are chosen to be YSOs.

Figure 8 shows the spatial distribution of both class I  and class
II sources. From  Figure 8, we note that class I and class II
sources are not symmetrically distributed in the whole selected
region, and are mostly concentrated in the north and southeast
around  SNR G59.5+0.1, which is similar to the results of Billot et
al. (\cite{billot}). Regarding the geometric distribution of class I
and class II sources, we can plot the map of the star surface
density. Because class I sources is younger than class II sources
and the spatial distribution of class II sources is similar to that
for class I sources, we only plot the map of the star surface
density for class I sources. This map was obtained by counting all
class I sources with a detection in the 3.6 $\mu$m, 4.5 $\mu$m, 5.8
$\mu$m, and 8.0 $\mu$m bands in squares of $4'\times 4'$, as shown
in Figure 9 . From Figure 9 we can see that there are clear signs of
clustering toward the region where SNR G59.5+0.1 is close
approximately to the surrounding ISM. The existence of class I
sources may also indicate star formation activity.

   \begin{figure}
  \vspace{-29mm}
\centering
\includegraphics[angle=0,scale=.535]{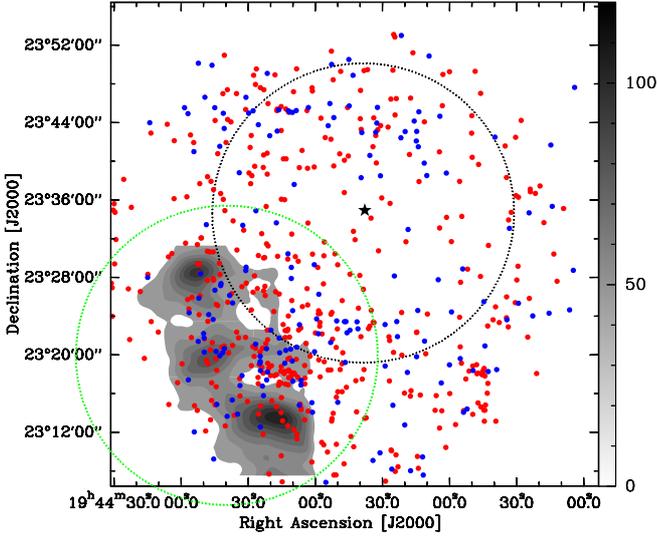}
 \vspace{-1mm}
\caption{Positions of  Class I and II  sources relative to the
$^{12}$CO $J=2-1$ MCs (grey scale) around SNR G59.5+0.1. The Class I
sources are labeled as the blue dots; The Class II sources are
labeled as the red  dots. The black and green dashed circles
represent SNR G59.5+0.1 and open cluster NGC6823, respectively.}
   \end{figure}

   \begin{figure}
   \vspace{3mm}
   \centering
   \includegraphics[angle=270,scale=.48]{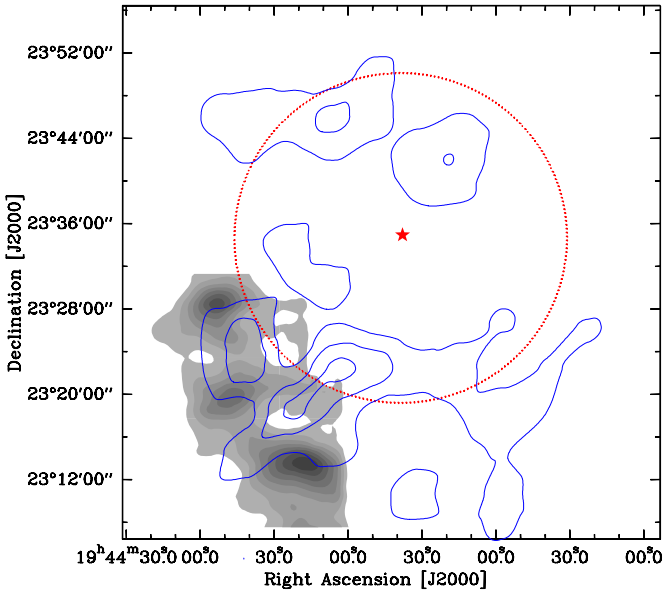}
   \caption{ Stellar-surface density map of
Class I candidates are superimposed on $^{12}$CO $J=2-1$ MCs (grey
scale) around SNR G59.5+0.1. Contours range from 2 to 10 stars
(4arcmin)$^{-2}$ in steps of 4 stars (4arcmin$)^{-2}$. 1$\sigma$ is
0.7 (4arcmin)$^{-2}$ (background stars).}
   \end{figure}

\section{Discussion}
\subsection{SNR G59.5+0.1 and MCs}
Because of the high confusion level in the surrounding of Sh2-86
structured emission, both Reach et al. (\cite{Reach}) and Billot et
al. (\cite{billot}) did not detect SNR G59.5+0.1. Similarly, we
could not detected the SNR on the 8.28 $\mu$m image, but detected
the 8.28 $\mu$m emission which present the half-shell structure
around the SNR. The half-shell emission may be produced by SNR
G59.5+0.1 and/or the stellar winds of its progenitor. We have
performed the submillimeter/millimeter observations in CO lines
toward the southeast of SNR G59.5+0.1.  The $^{12}$CO $J=2-1$
molecular emission is coincident very well with the 8.28 $\mu$m
emission, which covers the whole size of NGC 6823. In addition, the
$^{12}$CO $J=2-1$ molecular emission shows the arc-like morphology.
We identified three clumps in the molecular arc. Three bipolar
outflows are detected in these clumps around SNR G59.5+0.1 for the
first time. The line profile of $^{12}$CO $J=2-1$ in clump A appear
to be broadened, which may be not only caused by the outflows, but
also by shock from SNR G59.5+0.1. The maximum value of R$_{I_{\rm
CO(3-2)}/I_{\rm CO(2-1)}}$ is 1.6 in the molecular arc, which is
higher than that in the previous measurement of individual Galactic
MCs. High R$_{I_{\rm CO(3-2)}/I_{\rm CO(2-1)}}$ is suggested as a
signature of the SNR-MC interaction system (Jiang et al.
\cite{Jiang}; Xu et al. \cite{xua}). Together, these observations
strongly indicate that the MCs are interacting with G59.5+0.1.
SNR G59.5+0.1 just start to interact with the surrounding MCs, so
the MCs have the R$_{I_{\rm CO(3-2)}/I_{\rm CO(2-1)}}$ gradient
along the shock direction. We suggest that the distribution of
R$_{I_{\rm CO(3-2)}/I_{\rm CO(2-1)}}$ value is a good signature of
the SNR-MCs interaction system. The integrated CO intensity ratios
(R$_{I_{\rm CO(3-2)}/I_{\rm CO(2-1)}}$) may be used to study other
triggering mechanism of star formation, such as the shocks of
cloud-cloud collision, HII regions and galactic density waves. From
Table 2, the total mass of MCs associated with SNR G59.5+0.1 is
$1.1\times10^{4}\rm~ M_{\odot}$.

\subsection{ Recent star formation in the MCs }
To summarize the results from investigating the spatial distribution
of YSOs, we clearly see that the distribution of YSOs (class I and
class II sources) is clustered around the border of SNR G59.5+0.1.
The best correlation is found with the CO molecular arc in the
southeast where most of YSOs may belong to open cluster NGC6823. It
is unlikely that they all are just foreground and background stars.
It is more likely that those YSOs are physically associated with the
interacting regions between G59.5+0.1 and MCs. Riaz et al.
(\cite{Riaz}) suggested that NGC6823 have possibly experienced a
recent burst of star formation, which may be caused by a supernova
explosion of massive O star. Also, these YSOs distribution is
clustered together and shows a shell-like structure around
G59.5+0.1. SNR G59.5+0.1 is close approximately to NGC6823. We
suggest that SNR G59.5+0.1 possibly have triggered these YSOs
formation and star formation in this cluster. Class I sources occur
in a period on the order of $\sim$$10^{5}$ yr, while the age of
class II sources is $\sim$$10^{6}$ yr (Andr\'{e} \& montmerle 1994).

In an inhomogeneous medium, an SNR can be undergoing different
evolutionary stages in different places at the same time. If an SNR
is in radiative expansion phase, the radius of an SNR is $\sim$30 pc
(Heiles \cite{Heiles}). Here the radius of  SNR G59.5+0.1 is
$\sim$10 pc, the SNR has not yet been achieved the radiative phase,
but in adiabatic expansion. The age of G59.5+0.1 given by Sedov
equation (Reynoso \& Mangum  \cite{Reynoso}) in units of $10^{4}\rm
yr$:
\begin{equation}
\mathit{t}_{\rm S}=(R_{\rm
S}/13.6)^{5/2}(\frac{n_{0}}{E_{51}})^{1/2} .
\end{equation}
where $E_{51}$ is the remnant energy, we adopt a canonical value of
$10^{51}$ ergs. $R_{\rm S}$ is the remnant radius, while $\rm n_{0}$
is the the ambient number density in cm$^{3}$, which is determined
by Table 2. Then we obtain that the age of G59.5+0.1 is
8.6$\times$$10^{4}$ yr. The age of G59.5+0.1 are shorter than that
of these YSOs, which is 10$^{5}$ to 10$^{6}$ yr. Hence the formation
of the YSOs may not be triggered by the shock of SNR G59.5+0.1.
Because SNR G54.4-0.3, SNR G24.7+0.6 and SNR IC443 are not old
enough, Junkes et al. (\cite{Junkes}), Petriella et al
(\cite{Petriella10b}), and Xu et al. (\cite{xua}) suggested that the
YSOs around the remnants may not be triggered by the remnants, but
triggered by the stellar winds of the remnants progenitors.  SNR
G59.5+0.1 is located in the direction of Vulpecula OB at the
position ($l, b$) = (59.58, 0.12), hence  the spectral type of SNR
G59.5+0.1 progenitor may be O or B. In the course of high-mass stars
evolution the stellar winds go through three phases (Dwarkadas
\cite{Dwarkadas}), which last for about 4.6$\times$$10^{6}$ yr. Such
periods of stellar winds are sufficient to form YSOs.  We conclude
that the formation of the YSOs may be triggered by the stellar winds
from the high-mass progenitor of G59.5+0.1.

Outflow is a strong evidence of the earlier star forming activity.
We detected three outflows around SNR G59.5+0.1. The dynamic
timescale of each outflow is given by equation t = 9.78 $\times$
$10^{5}$R/V (yr), where R in pc is the outflow size defined by the
average of the radius of the blueshifted and redshifted lobes, V in
km $\rm s^{-1}$ is the maximum flow velocity relative to the cloud
systemic velocity. The obtained average dynamical timescales of
outflow A-C are $\sim$7.4$\times$$10^{3}$ yr,
$\sim$5.3$\times$$10^{3}$ yr, and $\sim$5.2$\times$$10^{3}$ yr,
suggesting that there are three Class 0 protostars ($\leq$ $10^{4}$
yr). These Class 0 protostars are distributed around SNR G59.5+0.1.
SNR G59.5+0.1 with a age of 8.6$\times$$10^{4}$ yr may trigger the
formation of these Class 0 protostars. The observations with higher
spatial resolutions are needed to further analyze these protostars.

\section{Conclusions}

We have presented the $^{12}$CO $J=2-1 $, $^{12}$CO $J=3-2$, and
$^{13}$CO $J=2-1$ molecular and infrared observations towards SNR
G59.5+0.1. These results can be summarized as follows:
 \begin{enumerate}
      \item The arc-like morphological association with SNR G59.5+0.1,
the broadened emission lines, and the high integrated CO line
intensity ratio (R$_{I_{\rm CO(3-2)}/I_{\rm CO(2-1)}}$)($\sim1.58$)
suggest that these cloud clumps are interacting with G59.5+0.1. The
age of the SNR is  8.6 $\times$ $10^{4}$ yr. The whole molecular gas
in the southeast have a line intensity ratio gradient along the
direction of shock, implying that shocks have driven into cloud
clumps. We suggest that high R$_{I_{\rm CO(3-2)}/I_{\rm CO(2-1)}}$
is identified as one good signature to study the triggering
mechanism of star formation, such as the shocks of cloud-cloud
collision, HII regions, and Galactic density waves. The cloud clumps
have the total mass of $1.1\times10^{4}\rm~M_{\odot}$.
      \item  The selected young stellar objects (YSOs)
(including class I and  class II sources) are concentrated and
grouped around the interacting regions. It provides strong signpost
for ongoing star formation. Comparing the age of G59.5+0.1 and the
timescales of the stellar winds of G59.5+0.1 progenitor with the
characteristic star-formation timescales, we conclude that the YSOs
may not be triggered by SNR G59.5+0.1, but triggered by the stellar
winds of G59.5+0.1 progenitor.
      \item  Three young outflows are detected in the clumps around SNR
G59.5+0.1. It may be the first SNR around which find such more
outflows in all the Galaxy SNRs. Taking into account the age of  SNR
G59.5+0.1 and class 0 source, we find that SNR G59.5+0.1 may trigger
the formation of these Class 0 sources. It also may provide us a
direct evidence for star formation triggered by SNR. The results of
the present work confirm the sequential star formation on the basis
of the different ages of star formation associated with SNR
G59.5+0.1.
   \end{enumerate}

\begin{acknowledgements}
We would like to thank Dr. Sheng-Li Qin for his help on data
acquirement and discussion. We also thank the anonymous referee for
his/her constructive comments and suggestions that greatly improved
the content and presentation of this paper. Jin-Long Xu's research
is in part supported by 2011 Ministry of Education doctoral academic
prize. Supported by the young Researcher Grant of National
Astronomical Observations, Chinese Academy of Sciences.
\end{acknowledgements}


\begin{thebibliography}{}
\bibitem[2004]{Allen} Allen, L. E., Calvet, N., D'Alessio, P., 2004, ApJS, 154, 363
\bibitem[1994]{Andre}  Andr\'{e}, P., \&  Montmerle, T.,  1994, \apj, 420,
837
\bibitem[1989]{Andres} Anders, E., \& Grevesse, N. 1989, Geochim. Cosmochim. Acta, 53, 197
\bibitem[2010]{billot} Billot, N., Noriega-Crespo, A., Carey, S., Guieu, S., Shenoy, S., Paladini, R., \& Latter, W., 2010, \apj, 712, 797
\bibitem[2006]{Byun} Byun, D.-Y., Koo, B.-C., Tatematsu, K., Sunada, K. 2006, \apj, 637,
283
\bibitem[1986]{Casoli} Casoli, F., Combes, F., Dupraz, C., Gerin, M., \& Boulanger, F., 1986, A\&A, 169, 281
\bibitem[1978]{Dickman} Dickman, R. L., 1978, \apj, 37, 407
\bibitem[2007]{Dwarkadas} Dwarkadas, V. V.,  2007, \apj, 667, 226
\bibitem[1996]{Frail} Frail, D. A., Goss, W. M., Reynoso, E. M., Giacani, E. B., Green, A.
J., \& Otrupcek, R. 1996, AJ, 111, 1651
\bibitem[2004]{Fazio} Fazio, G. G., Hora, J. L., Allen, L. E., et al. 2004, ApJS, 154, 10
\bibitem[1989]{Fich} Fich, M., Blitz, L., \& Stark, A. A., 1989, \apj, 342, 272
\bibitem[1991]{Garden} Garden, R. P., Hayashi, M., Hasegawa, T., et al.,  1991, \apj, 374, 540
\bibitem[1997]{Green}  Green, D. A., Frail, D. A., Goss, W. M., \& Otrupcek, R. 1997, AJ,
114, 2058 Hartmann, D., \& Burton, W. B. 1997, Atlas of Galactic
Neutral Hydrogen (Cambridge: Cambridge Univ. Press)
\bibitem[1993]{guesten} Guesten, R., Serabyn, E., Kasemann, C., et al., 1993, \apj, 402, 537
\bibitem[1992]{guetter} Guetter, H. H. 1992, \aj, 103, 179
\bibitem[1964]{Heiles} Heiles, C., 1964, \apj, 140, 470
\bibitem[2009]{Hewitt}  Hewitt, J. W., \& Yusef-Zadeh, F. 2009, \apj, 694,
L16
\bibitem[1986]{Huang}  Huang, Y.-L., \& Thaddeus, P. 1986, \apj, 309,
804
\bibitem[2010]{Jiang} Jiang, B., Chen, Y., Wang, J. Z. et al., 2010,
\apj, 712, 1147
\bibitem[1992]{Junkes} Junkes, N., F\"{u}rst, E., \& Reich, W., 1992, A\&A, 261, 289
\bibitem[2005]{Kharchenko} Kharchenko, N. V., Piskunov, A. E., R\"{o}er, S., Schilbach, E., \& Scholz, R.-D. 2005, A\&A, 438,
1163
\bibitem[1980]{Lang}  Lang, K. R.  1980, Astrophysical Formulae (Berlin:
Springer-Verlag), 157
\bibitem[2009]{Leao} Leao, M. R. M., de Gouveia Dal Pino, E. M., et al., 2009, MNRAS, 394,
157
\bibitem[2006]{Melioli} Melioli, C., de Gouveia Dal Pino, E. M., et al., 2006, MNRAS, 373,
811
\bibitem[2009]{Parons} Parons, S., Ortega, M. E., rubio, M., \& Dubner, G.,  2009, A\&A, 498,
445.
\bibitem[2010a]{Petriella10a} Petriella, A., Paron, S., \& Giacani, E., 2010,
A\&A, 513, A44
\bibitem[2011]{Petriella10b} Petriella, A., Paron, S., \& Giacani, E., 2011, Bolet\'{\i}n de la Asociaci\'{o}n Argentina de
Astronom\'{\i}a, 153, 221
\bibitem[2006]{Reach} Reach, W. T., et al. 2006, \aj, 131, 1479
\bibitem[2001]{Reynoso} Reynoso, E. M., \& Mangum, J. G., 2001, \aj, 121, 347
\bibitem[2012]{Riaz}  Riaz, B.,  Mart\'{\i}n, E. L.,  Tata R., et al., 2012, MNRAS, 419,
1887
\bibitem[1993]{sanders} Sanders, D. B., Scoville, N. Z., Tilanus, R. P. J.,  et al.,  1993, in Back to the Galaxy, et. S. S. Holt \& F. Verter (New York: AIP), 311
\bibitem[1998]{Seta}  Seta, M., et al. 1998, \apj, 505, 286
\bibitem[2008]{Sun}  Sun, K., Ossenkopf, V., Kramer, C., et al., 2008, A\&A, 489, 207
\bibitem[1992]{Taylor} Taylor, A. R., Wallace, B. J., \& Goss, W. M., 1992, \aj, 103, 931
\bibitem[1997]{wilson} Wilson, C. D., Walker, C. E., \& Thornley, M. D., 1997, \apj, 483, 210
\bibitem[1998]{Wilner} Wilner, D. J., Reynolds, S. P., \& Moett, D. A. 1998, \aj, 115,
247
\bibitem[2010]{xu} Xu, J. L.,  \& Wang, J. J., 2010, RAA, 2, 151
\bibitem[2011a]{xua} Xu, J. L., Wang, J. J., \& Miller, M., 2011a, \apj, 721, 81
\bibitem[2011b]{xub} Xu, J. L., Wang, J. J., \& Miller, M., 2011b, RAA, 11, 537
\bibitem[2005]{xuj} Xu, J. W., Zhang, X. Z., \& Han, J. L., 2005, Chin. J. Astron. Astrophys., 5, 165





\end{thebibliography}
\end{document}